\documentstyle[prd,floats,aps]{revtex}
\input psfig
\pssilent
\def\ut#1{#1\llap{\lower2ex\hbox{$\widetilde{\hphantom{#1}}$}}}
\begin{document}
\input psfig \pssilent 
\title{Canonical quantum gravity in the Vassiliev invariants arena:\\
II. Constraints, habitats and consistency of the constraint algebra}
\author{
Cayetano Di Bartolo$^{1}$
Rodolfo Gambini$^{2}$, Jorge Griego$^2$, Jorge Pullin$^3$}
\address{1. Departamento de F\'{\i}sica, Universidad Sim\'on Bol\'{\i}var,\\
Aptdo. 89000, Caracas 1080-A, Venezuela.}
\address{2. Instituto de F\'{\i}sica, Facultad de Ciencias, 
Igu\'a 4225, esq. Mataojo, Montevideo, Uruguay.}
\address{ 3. Center for Gravitational Physics and Geometry, 
Department of Physics,\\ The
Pennsylvania State University, 104 Davey Lab, University Park, PA
16802.} 

\maketitle
\begin{abstract}
In a companion paper we introduced a kinematical arena for the discussion
of the constraints of canonical quantum gravity in the spin network
representation based on Vassiliev invariants. In this paper we
introduce the Hamiltonian constraint, extend the space of states to
non-diffeomorphism invariant ``habitats'' and check that the off-shell
quantum constraint commutator algebra reproduces the classical Poisson
algebra of constraints of general relativity without anomalies. One
can  therefore consider the resulting set of constraints and space of
states as a consistent theory of canonical quantum gravity.
\end{abstract}
\vspace{-8.5cm} 
\begin{flushright}
\baselineskip=15pt
CGPG-99/11-1  \\
gr-qc/9911010\\
\end{flushright}
\vspace{6cm}

\section{Introduction}

\subsection{Preliminaries}
In our previous paper \cite{DiGaGrPu99}, we extended the notion of
Vassiliev invariant and other invariants associated with Chern--Simons
theory to the context of spin networks \cite{RoSm95}. We showed that
these invariants are loop differentiable in the sense of distributions
and we defined an infinitesimal generator of diffeomorphisms in terms
of the loop derivative. The generator correctly annihilates
diffeomorphism invariant states.  In this paper we will introduce a
Hamiltonian constraint based on the loop derivative. We will limit the
discussion to invariants of trivalent vertices and will only
concentrate on the ``Euclidean'' part of the Hamiltonian constraint in
the sense of Barbero \cite{Ba} and Thiemann \cite{QSD1}. Limiting
ourselves to trivalent intersections is clearly unphysical since the
volume operator identically vanishes on such states. However, the
essence of most calculations is already present with trivalent
intersections and the calculational difficulty is significantly lower
than with four or higher valent intersections. We will come back to
this issue in the discussion section where we will highlight which
results are quite plausibly going to survive when one goes to higher
valent intersections. We will then extend the space of invariants
through the construction of ``habitats'' such that the resulting
states are not diffeomorphism invariant. Based on these states we will
check the consistency of the quantum commutator algebra of
diffeomorphisms and Hamiltonian. We will show that there are no
anomalies, therefore constituting a consistent theory of canonical
quantum gravity. 

\subsection{Strategy}

We will start in the next section by introducing the Hamiltonian
constraint of quantum gravity in terms of the loop derivative. This
derivation will be ``generic'' in the sense that we will motivate it
through formal manipulations of the loop transform and therefore will
not make specific assumptions about the quantum state in the
connection representation on which the operator is acting upon.
Without being more precise about the space of states upon which it
acts, we cannot guarantee that the operator is well defined.  We will
show that on several spaces related to the Vassiliev invariants the
operator is indeed well defined.  The results obtained will be later
applied in specific ``habitats'' in which we will show that the
operators exist and we can compute correctly the commutators. This we
will discuss in section III. The need to introduce new
(non-diffeomorphism-invariant) ``habitats''
is based  on the fact that the spaces we have considered in the
companion paper are all diffeomorphism invariant and therefore one
cannot explore in a non-trivial way commutators involving the
diffeomorphisms. We will introduce an explicit example of such a
habitat in detail, the ``functions with marked vertices''. In the end
we will see that we will be able to recover the classical Poisson
algebra at the level of quantum commutators, but there will be subtle
issues involved in the definition of the right hand side of the
commutator of two Hamiltonians, which we will discuss in detail. 
The last section will be devoted to discussing the implications of the
level of consistency achieved by the theory. 

\section{The Hamiltonian constraint}
\label{hamcons}
We wish to introduce a quantum version of the Hamiltonian constraint
of canonical general relativity. Following Thiemann \cite{QSD1}, one
can construct this operator introducing the real version of the
Ashtekar variables first discussed by Barbero \cite{Ba}. In this
formulation, the canonical pair consists of a set of (densitized)
triads $\tilde{E}^a_i$ and as conjugate variables real-valued $SU(2)$
connections $A_a^i$. The Hamiltonian constraint of canonical, real,
Lorentzian general relativity in this framework consists of two terms,
\begin{equation}
H(N)= H_E(N)+ 
\int d^3x N(x) {4 \over G^3}\{A_a^i(x),K\}
\{A_b^j(x),K\} \{A_c^k(x),V\} \epsilon_{ijk} \tilde{\epsilon}^{abc}\;,
\end{equation}
where 
\begin{eqnarray}
H_E(N)&=&\int d^3x N(x) 
{2 \over G} \{A_a^i(x),V\} F_{bc}^i(x) \tilde{\epsilon}^{abc}\label{hame}
\;,\\
K &=& -{1 \over G}\{V, H_E(1)\}\label{K}\;,\\
V &=& \int d^3x 
\sqrt{\ut{\epsilon}\raise2ex\hbox{$$}\,{}_{abc} 
\tilde{E}^a_i(x) \tilde{E}^b_j(x) \tilde{E}^c_k(x)
\epsilon^{ijk}}\;.
\end{eqnarray}

In these expressions, $G$ is the Newton constant and in (\ref{K})
$H_E(1)$ means that the smearing function $N(x)$ is unity. This
remarkable formulation implies that the Hamiltonian constraint is
composed by two terms.  The first one, $H_E$, coincides with the
Hamiltonian constraint one would obtain in a canonical formulation of
general relativity on a manifold of Euclidean signature. The
Lorentzian theory is therefore attained through the addition of an
extra term. Both terms are given as expressions involving the Poisson
bracket of a connection with the volume of the space $V$ and with the
function $K$. The latter can be in turn obtained as a Poisson bracket
of the Euclidean part of the Hamiltonian constraint with the volume.

In this paper we will only concentrate on realizing at a quantum
mechanical level the Euclidean part of the Hamiltonian. If the reader
wishes, we are considering canonical quantum gravity in the Euclidean
context. However, it should be noticed that the additional term in the
Hamiltonian is obtained by successive commutators of the operator we
will consider, with the volume operator. This suggests that one could
generalize our construction to the Lorentzian case rather
straightforwardly, but this has not been studied in detail. The work
of Thiemann also shows that similar structures arise if one couples
the theory to matter. We will in this paper only concentrate on the
vacuum case but again one does not see a priori impediments to extend
our treatment to general relativity coupled to matter.

To implement the Hamiltonian constraint we will partially use the same
procedure proposed by Thiemann. We will not describe it again in full
detail here, we just list some of the salient features. One starts by
writing the classical expression (\ref{hame}) as
\begin{equation}
H(N)={2 \over G}\lim_{\epsilon\rightarrow 0}
\int d^3x \int d^3y N(y) \epsilon^{abc} {\rm Tr}\left[
F_{ab}(y) \{A_c(x),V\}\right] f_\epsilon(y,x)
\end{equation}
where $f_\epsilon(y,x)$ is a regularization of the Dirac
delta function, i.e. $\lim_{\epsilon\rightarrow 0} f_\epsilon(y,x)=
\delta^3(y,x)$. Strictly speaking, this expression is only gauge
invariant in the limit, so to preserve gauge invariance in the
regularization procedure we will ``join'' the $F_{ab}$ with the
connection using an infinitesimal piece of holonomy $h(\pi_x^y)$ along
an arbitrary path $\pi$ going from $x$ to $y$, which tends to the
null path in the limit $\epsilon\rightarrow 0$. To unclutter the
notation we will therefore denote $F_{ab}(\pi_x^y) \equiv
h(\pi_x^y) F_{ab}(y) h(\pi_y^x)$, 
\begin{equation}
H(N)={2 \over G}\lim_{\epsilon\rightarrow 0}
\int d^3x \int d^3y N(y) \epsilon^{abc} {\rm Tr}\left[
F_{ab}(\pi_x^y) \{A_c(x),V\}\right] f_\epsilon(y,x) \label{uno}
\end{equation}

In order to prepare the classical expression for quantization, we
triangulate the space-like hypersurface $\Sigma$ in terms of
elementary tetrahedra $\Delta$.  The triangulation has the following
properties: we select a finite set of distinct points of $\Sigma$,
denoted as $\{v\}$. At each of these points we choose three
independent directions $(\hat{u}_1, \hat{u}_2, \hat{u}_3)$ and
construct the eight tetrahedra with vertex $v$ and edges $(\pm u_1,
\pm u_2, \pm u_3)$, with $u_i=\epsilon \hat{u}_i$. The eight
tetrahedra saturating $v$ define a closed region $\Box_v$ of length
scale $\epsilon$. The remaining region $\Sigma - \cup_{v}\Box_v$ is
triangulated by arbitrary tetrahedra $\Delta'$. The motivation for
this peculiar discretization is that when we promote the classical
expression to a quantum operator, we will adapt the triangulation to
the spin network of the state in question by choosing the points $v$
to coincide with the vertices of the spin network. In terms of this
triangulation of space we can write equation (\ref{uno}) in the form,
\begin{equation}
H(N) = {2 \over G}\lim_{\Diamond\rightarrow 0} \int d^3y
\sum_{\Diamond\in\{\Box_v,\Delta'\}} {\cal V}_{\Diamond}
N(y) \epsilon^{abc} {\rm Tr}\left[ F_{ab}(\pi_{v_\Diamond}^y) \{
A_c(v_\Diamond), V\}\right] f_\Diamond(y)\,,\label{hamt}
\end{equation}
where the regions $\Diamond$ indicate either a box $\Box_v$ or a
tetrahedron $\Delta'$, and $v_{\Diamond}$ is any point interior to
$\Diamond$. Conventionally we will choose for the boxes
$v_{\Box_v}=v$, and for the tetrahedra $v_{\Delta'}$ will represent one
of its vertices. To discretize the integral we have introduced the
volume of each region ${\cal V}_{\Diamond}$,
\begin{equation}
{\cal V}_{\Diamond} \epsilon^{abc} = \frac{\alpha_{\Diamond}}{6}
 \epsilon^{ijk} u^a_i u^b_j u^c_k\,,\label{vol}
\end{equation}
where $\alpha_{\Diamond}=8$ if $\Diamond=\Box$, and
$\alpha_{\Diamond}=1$ if $\Diamond=\Delta'$. In this last case, the
$u_i$'s represent the edges of the tetrahedra $\Delta'$ adjacent to
$v_{\Delta'}$. We also have adapted the regularization of the Dirac delta
function to the tetrahedral decomposition by defining,
\begin{equation}
f_{\Diamond}(y) = {\Theta_{\Diamond}(y) \over {\cal
V}_{\Diamond}}\,,
\label{epsi}
\end{equation}
where $\Theta_{\Diamond}(y)$ is one if $y\in\Diamond$ and zero
otherwise. We now replace in (\ref{hamt}) the $\epsilon^{abc}$
using formula (\ref{vol}) and we represent $u^c_k A_c(v_\Diamond)$
using a holonomy in the fundamental representation along the edge
$u_k$ of the triangulation,
\begin{equation}
H(N) =\lim_{\Diamond\rightarrow 0} \int d^3y
\sum_{\Diamond\in\{\Box_v,\Delta'\}} \frac{\alpha_{\Diamond}}{3G}
\,\epsilon_{ijk}\, u_i^a u_j^b\, N(y)\, {\rm
Tr}\left[ F_{ab}(\pi_{v_{\Diamond}}^y) h(u_k) \{
h^{-1}(u_k),V\} \right]f_\Diamond(y)\,, \label{classham}
\end{equation}
and we retraced the path described by the
holonomy to preserve gauge invariance. It should be noticed that
this retracing does not contribute to the expression at leading
order in $\epsilon$ (the scale parameter of the regions
$\Diamond$). We will now proceed to study the quantization of the
last expression. For that we need to ``adapt'' the triangulation
we introduced to a spin network state $s$ \cite{QSD1}. Let
$\{v_s\}$ be the set of vertices of the spin network and
let $(e_{1m}, e_{2m}, e_{3m})$ be the triples
of non-coplanar edges incident at $v_s$ ($m=1,\ldots,E(v_s)$, being
$E(v_s)$ the number of such triples). To adapt the triangulation
to the spin network state we perform two  operations: first we
identify the points $v$ of the boxes $\Box_v$ with the vertices
$v_s$ and second, given a triple of edges $(e_{1m}, e_{2m}, e_{3m})$
incident at $v_s$, we orientate the three unit vectors $(\hat{u}_1,
\hat{u}_2, \hat{u}_3)$ along the tangents of the edges at $v_s$.
Then we can write,
\begin{equation}
\epsilon \hat{u}^a_{im} =\int_{e_{im}} dw^a\,
\Theta_{\Box_{v_s}}(w)\,.\label{tangent}
\end{equation}
To simplify the notation we identify from now on $v_s\equiv v$. As
in the previous paper, we assume we are acting on a state given by
a loop transform,
\begin{equation}
\Psi(s) = \int DA \Psi[A] W_A[s]\,, \label{states}
\end{equation}
and we realize the Hamiltonian operator over the spin network
wavefunctions by promoting the classical expression (\ref{classham})
as an operator acting on the Wilson net appearing in the loop
transform, very much as we did in the companion paper when we
discussed the diffeomorphism operator,
\begin{eqnarray}
{H}(N) \Psi(s) &=& {8 \over 3G}\int DA \Psi(A)
\lim_{\Box\rightarrow 0} \int d^3y \sum_{v\in s} \sum_{m=1}^{E(v)}
\frac{\epsilon_{ijk}}{E(v)} \int_{e_{im}} dw^a\, \Theta_{\Box_v}(w)
\int_{e_{jm}} dw^b\, \Theta_{\Box_v}(u)
\times \\&&
\times
N(y)\, {\rm Tr}\left(F_{ab}(\pi_{v}^y) h(u_{km}) \left[
h^{-1}(u_{km}),{V}\right] \right)f_{\Box_v}(y) W_A(s)\,.\nonumber
\end{eqnarray}
A first observation is that the volume operator has non-vanishing
contributions only at the vertices of the spin net, so we replace
the sum in the Hamiltonian over all $\Diamond$'s by a sum over all
the vertices of the spin net. The second observation is that, for an
n-valent vertex, we have taken the average over all the non-coplanar
triples of edges associated with this vertex. Finally notice that
only one of the terms in the commutator contributes, that with the
volume operator on the left. The one with the volume operator on
the right vanishes since it is proportional to the trace of the Lie
algebra element ${\rm Tr}(F_{ab} h_{u_{km}} h^{-1}_{u_{km}})$. Let
us now consider a generic $n-$valent vertex and study the action of
the operator on the Wilson net. We assume that the holonomies are
all outgoing from $v$. Schematically, we are trying to represent
the action of ${\rm Tr}(F_{ab} h_{e} {V} h^{-1}_{e}) W_A(s)$,
with $e$ a generic edge of spin\footnote{We
use capital Latin letters to design
the ``spins'' of the edges of the spin network.}
 $J$ incident at $v$. An $n-$valent
vertex is characterized by an intertwiner that can be represented
through a vector $\vec{I}$ of $n-3$ of irreducible representations.
The factor $h^{-1}_{e}$ acts by adding a line of spin $1/2$
ingoing from the vertex $v$ and parallel to the line $e$,
\begin{equation}
\left(h_{e}\right)^D{}_C {V} \left(h^{-1}_{e}\right)^C{}_B
\,W_A\left(
\raisebox{-12.5mm}{\psfig{file=vern.eps,height=25mm}}\right) = 
\left(h_{e}\right)^D{}_C {V} \,
W_A\left(
\raisebox{-12.5mm}{\psfig{file=vernins.eps,height=25mm}}\,\right)\,.
\end{equation}
Notice that the insertion of the holonomy leaves a Wilson net that is
not gauge invariant anymore, we represent this by keeping the group
indices $B,C$ in the diagram. 

If one was acting on an $n-$valent vertex characterized by an
intertwiner $\vec{I}$, this leaves an $(n+1)-$valent vertex
characterized by intertwiners $\vec{I},J$; $J$ being in the diagram
the spin of the line connecting the original vertex to the point
(infinitesimally nearby) where the holonomy was inserted. Notice that
$J$ coincides with the spin of the original line $e$. We denote
that the line is infinitesimally close by the dashed circle. We now
act with the volume operator. For that we first have to reduce the
product of holonomies associated with $e$ to a superposition of
irreducible representations. This procedure defines a new
intertwiner for the vertex $v$ over which the action of the volume
operator is well-defined,
\begin{equation}
\left(h_{e}\right)^D{}_C {V} \,
W_A\left(
\raisebox{-12.5mm}{\psfig{file=vernins.eps,height=25mm}}\,\right)=
\left(h_{e}\right)^D{}_C {V} \,\sum_{K} (-1)^{2J}(2K+1)
W_A\left(
\raisebox{-12.5mm}{\psfig{file=verninsprprpr.eps,height=25mm}}\,\right)\,.
\end{equation}
The action of the volume operator can be represented by a matrix
that rearranges the intertwining of the vertex. This matrix
elements would depend also on the spins of the external edges of
the vertex (notice that $K$ is the color of one of this external
edges). Schematically,
\begin{eqnarray}
\left(h_{e}\right)^D{}_C {V}
W_A\left(
\raisebox{-12.5mm}{\psfig{file=verninsprprpr.eps,height=25mm}}\,\right)&=&
\left(h_{e}\right)^D{}_C
\sum_{\vec{I'},J'} V_{\vec{I},J}^{\vec{I'},J'}
W_A\left(
\raisebox{-12.5mm}{\psfig{file=verninsprpr.eps,height=25mm}}\,\right)
\nonumber\\
&=&
(-1)^{K+\frac{1}{2}+J}\sum_{\vec{I'},J'}
V_{\vec{I},J}^{\vec{I'},J'}
W_A\left(
\raisebox{-12.5mm}{\psfig{file=verninspr3.eps,height=25mm}}\,\right)   \,,
\end{eqnarray}
where in the last step we reduce the line of spin $1/2$ of the
holonomy using recoupling theory. One is left with a double
insertion at the vertex, which we can rearrange via recoupling,
since everything is happening infinitesimally close to the vertex.
We write the result in the following way,
\begin{eqnarray}
&&\left(h_{e}\right)^D{}_C {V} \left(h^{-1}_{e}\right)^C{}_B
\,W_A\left(
\raisebox{-12.5mm}{\psfig{file=vern.eps,height=25mm}}\right)\\
&&\hspace{2cm} =
\sum_{K,\vec{I'},J',K'} (-1)^{K+\frac{3}{2}-J}(2K+1)(2K'+1)
\left\{
\begin{array}{ccc}
J&J'&K'\\
{1\over 2}&{1\over 2}&K
\end{array}
\right\}
V_{\vec{I},J}^{\vec{I'},J'}
W_A\left(
\raisebox{-12.5mm}{\psfig{file=verninsret.eps,height=25mm}}\,\right).
\label{nval}\nonumber
\end{eqnarray}
The above expression has now to be contracted with $F_{ab}$ at the
open strands labeled with $B$ and $D$. One can generate $F_{ab}$ by
introducing an extension of the idea of loop derivative for spin
networks. The reason why one does not simply recover the ordinary loop
derivative is that the edge before and after the point of the
insertion of $F_{ab}$ are in general in different representations $J$
and $J'$. The ordinary loop derivative inserts an $F_{ab}$ without
changing the representation of the original line. 
This definition of loop derivative appeared as natural in the context
of loops, however in terms of spin networks one expects in general to
have a situation like the one we have here. One can use the same set
of constructions we did for the ordinary loop derivative of the
invariant $E(s,\kappa)$ to compute the action of this generalized loop
derivative\footnote{$\kappa$ is proportional to the inverse of the
coupling constant of Chern-Simons theory. For a definition of the
invariant $E(s,\kappa)$, see the companion paper, section
II.C.}. However, we will not pursue the exploration of this
generalized derivative in this paper. Instead we will concentrate on
the case of trivalent intersections. If the spin networks considered
have trivalent intersections, when we repeat the construction we did
above, the representations of the edges before and after the insertion
of $F_{ab}$ are the same. The operator we then obtain to represent
$F_{ab}$ corresponds in this case to the ordinary loop derivative. We
have now to consider the action of the volume operator over a
four-valent vertex with only three real edges \cite{BoDePiRo},
\begin{equation}
{V}\,
W_A\left(
\raisebox{-12.5mm}{\psfig{file=verninshalf.eps,height=25mm}}\,\right)=
\omega(L,M,K)\,
W_A\left(
\raisebox{-12.5mm}{\psfig{file=verninshalf.eps,height=25mm}}\,\right)\,,
\end{equation}
with \footnote{The result listed here differs from that given in
reference \cite{BoDePiRo}, because we are using a different
normalization in the definition of spin networks, details can be seen
in \cite{GaGrPu98}. Our convention agrees up to a numerical factor
with Thiemann's conventions \cite{Thivo}. }
\begin{equation}
\omega(L,M,K) =
\sqrt{\Lambda_{LMK}\sqrt{K(K+1)}\left|
\left\{ 
\begin{array}{ccc}
K&{1\over 2}&K+{1\over 2}\\
K-{1\over 2}&1&K
\end{array}
\right\}
\left\{ 
\begin{array}{ccc}
1&K-{1\over 2}&K+{1\over 2}\\
1&L&L\\
1&M&M
\end{array}
\right\}\right|}\,,
\end{equation}
where $\Lambda_{LMK}\equiv\Lambda_L\Lambda_M\Lambda_K$, and
$\Lambda_J=\sqrt{J(J+1)(2J+1)}$. Using this result, the matrix
elements of equation (\ref{nval}) are reduced to the following
simple expression for trivalent vertices ($\vec{I}=\vec{I'}\equiv0$
in this case),
\begin{equation}
V^{J'}_{J}(L,M,K)=\delta^{J'}_{J}\,\omega(L,M,K)\,,
\end{equation}
where $K=J\pm {1\over 2}$, and $L$ and $M$ are the other spins of
the edges incident at the trivalent vertex. Therefore the spin
of the link going from the intersection to the point of insertion
of $F_{ab}$ is unchanged through the action of the volume operator.
By inserting the $F_{ab}$ we obtain the final action of the
Hamiltonian constraint on trivalent vertices,

\begin{eqnarray}
{H}(N) \psi
\left(
\raisebox{-12.5mm}{\psfig{file=ver3.eps,height=25mm}}\right) &=& 
{8 \over 3G}
\lim_{\Box\rightarrow 0} \int d^3y \sum_{v\in s}
\epsilon_{ijk} \int_{e_i} dw^a \,\Theta_{\Box_v}(w) \int_{e_j} dt^b
\Theta_{\Box_v}(t)
N(y) f_{\Box_v}(y) \times \nonumber\\
&&\times \rho_k(\vec{J}_v)\,
\Delta^{(k)}_{ab}(\pi_{v}^y)
\psi
\left(
\raisebox{-12.5mm}{\psfig{file=ver3.eps,height=25mm}}\right)\,.
\label{hgen}
\end{eqnarray}
The superscript on the loop derivative denotes at which
of the incident edges at the intersection it adds the extra strand
$\pi$. Notice that since $k$ is summed by the Einstein convention, the
derivative ends up acting on all edges (we have abused of the
summation convention allowing the index $k$ to be repeated three times
in the last expression). The vector $\vec{J}_v=(J_1, J_2, J_3)$
labels the spins of the tree edges of the spin network incident at
$v$ ($J_i$ is the spin of the edge $e_i$), and the group factor
$\rho_k(\vec{J}_v)$ is given by the expression,
\begin{equation}
\rho_k(\vec{J}_v):=\sum_Q (-1)^{Q+\frac{1}{2}+J_k}
\,\sqrt{\frac{3}{2}}\,\frac{(2Q+1)}{\Lambda_{J_k}}\,
\left\{
\begin{array}{ccc}
J_k&J_k&1\\
{1\over 2}&{1\over 2}&Q
\end{array}
\right\}\, \omega(J_{k1}, J_{k2}, Q)\,,
\end{equation}
where the indices $(k, k1, k2)$ take cyclic values in the set $(1,
2, 3)$.

We have therefore a general expression for the Hamiltonian constraint
acting on any loop-differentiable wavefunction with support on
trivalent spin networks, that bears a relationship with a wavefunction
in the connection representation $\Psi(A)$ given by (\ref{states}). 

At this point it is worthwhile comparing this action of the
Hamiltonian constraint we just introduced with the original proposal
of a Hamiltonian constraint in the loop representation (doubly
densitized) in terms of loop derivative as introduced in
\cite{Ga91}. Such constraint was written in
\cite{GaGaPu} as,
\begin{equation}
{H}(\ut{N}) \Psi(\gamma)= {2 \over G} \lim_{\epsilon\rightarrow 0}
\epsilon \int d^3x \int_\gamma dy^a \int_\gamma dz^b \ut{N}(x)
f_\epsilon(y,z) \delta^3(y-z) \Delta_{ab}(\gamma_o^x) {O}_{y,z}
\Psi(\gamma) 
\end{equation}
The Hamiltonian acted on
functions of loops and the operator ${O}$ had the action of
re-routing one of the lobes of the partition of the loop determined by
the points $y,z$. We see two fundamental differences with the operator
we just introduced. The first one is the re-routing operator
${O}$. This operator arose in the loop representation to account
for the fact that states based on loops in the fundamental
representation do not diagonalize operators like the volume. In the
spin network context the operator is replaced by the group factors we
discussed. More important is the difference concerning density
weights. In the Hamiltonian in terms of loops we see that one obtains
a doubly densitized quantity by considering the product of a Dirac
delta times the regulator. To understand properly this difference, it
is worthwhile considering on which spaces of functions are these
operators meant to operate upon. In the Hamiltonian in terms of loops
one had in mind that the loop derivative was acting on functions of
loops such that the result was a smooth function. An example of such
functions would be holonomies built with a smooth connection. The
resulting expression would then consist of a smooth tensor, $F_{ab}$,
a Dirac delta integrated on three dimensional space and, in the limit,
a two dimensional Dirac delta given by $\lim_{\epsilon \rightarrow 0}
\epsilon f_\epsilon(y,z)$ integrated along the two one dimensional
integrals. The result is finite but is regularization dependent, since
the two dimensional Dirac delta has an inverse power of the three
dimensional volume in it that is not compensated.

If we now turn our attention to the Hamiltonian we introduced in this
paper, and consider its action on a state of the form 
$E(s,\kappa)$, we will see a different behavior. 
Given the action (companion paper: formulas 71 and 44) of
the loop derivative on $E(s,\kappa)$ we get,
\begin{eqnarray}
{H}(N) E
\left(
\raisebox{-12.5mm}{\psfig{file=ver3.eps,height=25mm}},\kappa\right) &=& 
-{8\kappa \over 3G}
\lim_{\Box\rightarrow 0} \int d^3y \sum_{v\in s}
\epsilon_{ijk} \int_{e_i} dw^a \Theta_{\Box_v}(w)\int_{e_j} dt^b
\Theta_{\Box_v}(t)
N(y) f_{\Box_v}(y) \times \nonumber\\
&&\times  \rho_k(\vec{J}_v) \epsilon_{abc}\int_{e_k} dz^c \delta^3(z,y)
E\left(
\raisebox{-12.5mm}{\psfig{file=ver3del.eps,height=25mm}},\kappa
\right), \label{hame2}
\end{eqnarray}
where the extra edge added by the loop derivative starts and ends
in the edge $e_k$ and as before, 
$(k, k1, k2)$ take cyclic values in the set $(1,
2, 3)$.
The reason for this is that if it ended in any
other edge incident on the vertex one would (in the limit
$\Box\rightarrow 0$) get zero for the action since one would have
an integral repeated over one of the edges contracted with
$\epsilon_{abc}$.

The action on $E(s)$ of the loop derivative that appears in the Hamiltonian
constraint, for trivalent vertices reduces therefore to a chord
diagram. If we evaluate the chord diagram using recoupling identities, we
finally get,
\begin{eqnarray}
{H}(N) E
\left(
\raisebox{-12.5mm}{\psfig{file=ver3.eps,height=25mm}},\kappa\right) &=& 
-{8\kappa \over 3G}
\lim_{\Box\rightarrow 0} \sum_{v\in s}\int d^3y 
\epsilon_{abc}\epsilon_{ijk} \int_{e_i} dw^a \Theta_{\Box_v}(w)
\int_{e_j} dt^b
\Theta_{\Box_v}(t) \times\\
&&\int_{e_k} dz^c \delta^3(z,y)
  N(y) f_{\Box_v}(y)\,\rho_k(\vec{J}_v)
\,J_k(J_k +1)\, E\left(
\raisebox{-12.5mm}{\psfig{file=ver3.eps,height=25mm}},\kappa
\right). \nonumber
\end{eqnarray}
At this point it is worthwhile continuing the comparison with the
doubly densitized case. If we evaluate the integrals involved in the
above expression we see that we have six one dimensional integrals and
two three dimensional Dirac delta functions. In the limit we therefore
have a finite result, that is defined independent of the background
structures introduced by the regulator $f_\Box(y)=\Theta_\Box(y)/{\cal
V}_\Box$. One first performs the integral in $y$, which fixes $y=z$ and
then one is left with three one dimensional integrals with three step
functions (two explicit, one present in the $f_{\Box_v}$). The result
of the integrals is ${\cal V}_{\Delta}$, which cancels the denominator
of the $f_\Box$ (${\cal V}_{\Box}=8{\cal V}_{\Delta}$) and one is left
with a finite result, that in the limit $\Box\rightarrow 0$ gives,
\begin{equation}\label{nu}
{H}(N) E
\left(
\raisebox{-12.5mm}{\psfig{file=ver3.eps,height=25mm}},\kappa\right) = 
-{\kappa \over G}
\sum_{v\in s} N(v) \, \rho(\vec{J}_v)\,
E\left(
\raisebox{-12.5mm}{\psfig{file=ver3.eps,height=25mm}},\kappa
\right), \nonumber
\end{equation}
where
\begin{equation}
\rho(\vec{J}_v):=4 \sum_{k=1}^{3}J_k(J_k +1)\rho_k(\vec{J}_v)\,,
\end{equation}
is the factor group associated with the vertex $v$.

We have introduced a quantum version of Thiemann's
singly-densitized classical Hamiltonian in terms of the loop
derivative that has the remarkable following properties:

a) It has a significant resemblance to the original doubly-densitized 
Hamiltonian  proposed in the loop representation \cite{Ga91}, which in
turn closed the constraint algebra formally \cite{GaGaPu}. We will
discuss the implications of this in the next section.

b) The Hamiltonian introduced is finite on the space of Vassiliev
invariants we introduced in the previous paper. This happens due to the
detailed form in which the loop derivative acts on these states. 

c) The Hamiltonian may be well defined on other spaces of states. For
instance, on the space of functions of spin nets obtained by
considering the Wilson nets along smooth holonomies (the original kind
of function that was first considered as loop differentiable), it is
straightforward that the Hamiltonian identically vanishes
(essentially, since the loop derivative is smooth, the integrals that
before gained contributions due to the distributional character of the
loop derivative now vanish).

d) The last two points show that something surprising is happening
here, in the sense that {\em we did not construct the Hamiltonian in
an ad-hoc way to obtain properties b) and c)}. That is, we introduced
a discretization of the singly-densitized Hamiltonian and it naturally
turned out to be well defined on the Vassiliev invariants and to
vanish on holonomies of smooth connections. This might be pointing to
a certain ``naturalness'' of the Vassiliev invariant space in the
context of quantum gravity. In spite of the a priori huge ambiguity in
the regularization of the singly-densitized Hamiltonian, it is
difficult to imagine a regularization involving the loop derivative
that would not be finite on the Vassiliev invariants. In addition,
we have not introduced in our construction any ad-hoc ``renormalization''
of the operator to obtain finiteness (as one had to do in the
doubly-densitized case).

We will now proceed to discuss the issue of the constraint algebra.

\section{Consistency of the constraint algebra: preliminaries}
The operators we introduced are written purely in terms of the loop
derivative and integrals along edges of the spin net. One can discuss
the successive action of such operators without making specific
reference to a space of states on which one is acting upon. Such a
calculation would be ``formal'' in the sense that terms arising in it
could fail to be well defined when acting on particular spaces of
states. This is the underlying philosophy of the kind of (unregulated)
formal calculations some of us undertook in reference
\cite{GaGaPu}. Such an approach can be pursued with the regulated
operators we introduce in this paper. The fact that the operators are
regulated and that they might act on the space of diffeomorphism
invariant states will imply certain departures from the specifics of
the calculations of reference \cite{GaGaPu}. If one pursues
calculations at this level of generality, one cannot expect to
reproduce the classical Poisson algebra at the level of quantum
commutators without making some assumptions about the distributional
behavior of the various quantities involved in the calculations. One
will recover terms that correspond to the classical ones but
generically there will be additional terms. When one particularizes to
a certain ``habitat'' of wavefunctions, these extra terms in many
cases will vanish. However, it is not excluded that on certain
habitats pathologies could appear. It would be an interesting
exercise to pursue the calculation in general and then
particularize to various ``habitats''. This, however, is a complex
and lengthy calculation in terms of loop calculus. We will not
attempt a calculation at this level of generality here. In this
paper we will concentrate on computing the constraint algebra on a
series of  habitats on which we recover the classical constraint
algebra at the level of quantum commutators. These habitats will
include as particular cases spaces of diffeomorphism invariant
functions on which the appropriate algebra will be recovered.

\subsection{Functions of spin nets with ``marked vertices''}
\label{marked}
A point to be addressed when studying the constraint algebra is that,
given the action of the Hamiltonian constraint we have introduced, if
one starts with a function of spin networks, when one acts with a
Hamiltonian constraint one ends with a more general object. This
can be explicitly seen for instance in equation (\ref{nu}).
What we see is that the result of the action of a Hamiltonian 
on a function of spin networks is a function of spin networks times a
``vertex function'', that is, a function dependent on the position
of the vertex where the Hamiltonian acts. If one wishes to compute
the constraint algebra, the action of the second operator in the 
calculation of the algebra therefore takes place on such a space of
functions. One can view these functions as simply more general 
functions of spin networks, since after all the information of 
the position of the vertices comes with a given spin network. 
This is possible, but one has to exhibit that dependence -of the
vertices on the spin networks- in an explicit enough way that
operators like the loop derivative have an appropriate action on
such a dependence. One is initially tempted to say that the loop
derivative simply ignores the extra factors, since they do not
appear to contain an explicit loop dependence. But the position of
the vertices depends on the edges of the spin network (more
precisely, a vertex is defined by the intersection point of at
least three edges of the net), and therefore the vertex function
should be affected by the loop derivative operator. To operate with
the loop derivative over this kind of functions, we need to make
more explicit the implicit functional dependence $v[e(s)]$ of the vertex
function. We limit the analysis to the case of
trivalent vertices. Given three arbitrary open paths $\gamma_i$
($i=1,2,3$) with a common origin $o$, we define the quantity,
\begin{equation}
R_{c}(M,\epsilon,\gamma):={1\over 2}
\epsilon_{cdf}\int_{\bar{\gamma_3}\circ \gamma_1} dx_1^d
\int_{\bar{\gamma_3}\circ \gamma_2} dx_2^f
(M(x_1)+M(x_2))
\frac{\Theta_{\epsilon}(x_1 -x_2)}{\epsilon^2},\label{pdelta}
\end{equation}
where $\gamma$ denotes generically the three curves
$\gamma_1,\gamma_2,\gamma_3$, $\Theta_{\epsilon}(y)$ is $1$ if
$|y|<\epsilon$ and zero otherwise, and $M(x)$ is a scalar function
defined on the manifold. It is assumed that the three paths are
oriented and we denote by the overbar the path in the reversed
orientation.  We also assume that, starting at $o$, the three curves
overlap in a finite segment and then they separate following disjoint
paths, as it is shown in the figure.
\begin{center}
\begin{figure}[t]
\centerline{\psfig{figure=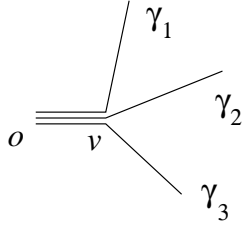,height=30mm}}
\caption{The three open paths.}
\end{figure}
\end{center}
We call $v$ the bifurcation point of the three curves. It is immediate
to show that,
\begin{equation}
R_{c}(M,\epsilon,\gamma)= M(v)
R_{c}(1,\epsilon,\gamma)+ O(\epsilon),
\end{equation}
and,
\begin{equation}
M_{\epsilon}(v):=
\frac{\vec{R}(M,\epsilon,\gamma)\cdot
\vec{R}(1,\epsilon,\gamma)}
{|R(1,\epsilon,\gamma)|^2}
= M(v)+O(\epsilon).\label{mv}
\end{equation}
The above expressions can be used to calculate the loop derivative
of a vertex function. Given a vertex of the net with edges $e_1,
e_2, e_3$, we identify the portion of the curve $\gamma_i$ going
form $v$ to the end point with the portion of the edge $e_i$ from
the vertex to some intermediate point. This automatically
guarantees that the point $v$ defined through the family of curves
$\gamma_i$ coincides with the chosen vertex of the graph. If the
end points of the edges $e_1, e_2, e_3$ lie outside of the ball of
radius $\epsilon$ centered at $v$, then one has the identity
$R_{c}(M,\epsilon,\gamma_1, \gamma_2,
\gamma_3)=R_{c}(M,\epsilon,e_1, e_2, e_3)\equiv
R_{c}(M,\epsilon,e)$. 

We can now proceed to compute explicitly the loop derivative. For any
function $F(x)$ we have (\cite{GaPubook} chapter 1),
\begin{equation}
\Delta^{(e)}_{ab}(\pi_v^z)\,\int_{e} dx^c F(x) =
\delta^c_{[b}\partial_{a]z} F(z).\label{loopder}
\end{equation}
Using this result it is straightforward to calculate the loop
derivative of $R_{c}(M,\epsilon,e)$,
\begin{eqnarray}
\Delta^{(e_1)}_{ab}(\pi_v^z)\,R_{c}(M,\epsilon,e) &=&
\frac{\epsilon_{cdf}}{2\epsilon^2} \int_{\bar{e_3}\circ e_2} dy^f
\left\{ \delta^d_{[b}\partial_{a]z} M(z) \Theta_{\epsilon}(z-y)\right.
\nonumber\\&&\left.+
(M(z)+M(y)) \delta(\epsilon-|z-y|)
\frac{\delta^d_{[b}(y-z)_{a]}}{|z-y|}\right\}.\label{derR}
\end{eqnarray}
The loop derivative of $M_{\epsilon}(v)$ is evaluated using the above
result and the Leibniz' rule.

Now that we know how to compute the loop derivative of a vertex function, we
can consider the construction of a ``habitat''
of functions of spin networks with marked points of the definite
following kind:
\begin{equation}
\psi_\epsilon(s,M,\Omega) =
\sum_{v\in s} M_\epsilon(v) \Omega(\vec{J}_v) \psi(s)\equiv
O_{\epsilon}(M,\Omega) \psi(s)\,
\end{equation}
where $\Omega(\vec{J}_v)$ is a group factor depending on the spins
of the edges incident at $v$. This type of dependence on the vertex
is precisely the one that appears in the action of the
Hamiltonian constraint (\ref{hame2}). In the last identity, we
express the relationship between $\psi_\epsilon(s,M,\Omega)$ and
$\psi(s)$ through an operator $O_{\epsilon}(M,\Omega)$. It is clear
that, in the limit $\epsilon\rightarrow 0$, one gets as a result a
function of the spin network times a function of the vertices and
incoming spins,
\begin{equation}
\psi(s,M,\Omega)\equiv O(M,\Omega)\psi(s):=
\lim_{\epsilon\rightarrow 0} \psi_\epsilon(s,M,\Omega) =
\sum_{v\in s} M(v) \Omega(\vec{J}_v) \psi(s)\,.\label{psiM}
\end{equation}
To define the action of the constraints on these kinds of functions
we will study the action of the constraints before taking the limit 
$\epsilon\rightarrow 0$ in these expressions in a suitable way. Let
us start by considering the diffeomorphism constraint,
\begin{eqnarray}
C(\vec{N}) \psi(s,M,\Omega) &=& \lim_{\{\epsilon, \epsilon'\}\rightarrow 0}
C_{\epsilon'}(\vec{N})O_{\epsilon}(M,\Omega) \psi(s)\nonumber\\
&=&\lim_{\{\epsilon, \epsilon'\}\rightarrow 0} \left[
\{C_{\epsilon'}(\vec{N}) O_{\epsilon}(M,\Omega)\} \psi(s)
+O_{\epsilon}(M,\Omega)C_{\epsilon'}(\vec{N})  \psi(s)\right]\,.
\end{eqnarray}
In the second term the diffeomorphism acts on $\psi(s)$ leaving all
the factors unaffected. This term is immediate to evaluate (and in
the case of diffeomorphism invariant $\psi(s)$'s vanishes), so we
will not discuss it further. In order to analyze the first
contribution it is convenient to make the following intermediate
calculation,
\begin{eqnarray}
C_{\epsilon'}(\vec{N})\,R_{c}(M,\epsilon,e)&=&
\sum_{e_m} \int d^3z \int_{e_m} dx^b
{N^a(x)+N^a(z)\over 2} f_{\epsilon'}(x,z)
\Delta_{ab}^{(m)}(\pi_x^z)\,R_{c}(M,\epsilon,e)\\
&=&\frac{\epsilon_{cdf}}{4\epsilon^2} \int_{\bar{e_3}\circ e_2} dy^f
\int_{e_1} dx^b \int d^3 z (N^a(x)+N^a(z)) f_{\epsilon'}(x,z)
\left\{ \delta^d_{[b}\partial_{a]z} M(z) \Theta_{\epsilon}(z-y)\right.
\nonumber\\
&&\left.+(M(z)+M(y))
\delta(\epsilon-|z-y|)\frac{\delta^d_{[b}(y-z)_{a]}}{|z-y|}\right\}+
\mbox{\footnotesize{cyclic permutations in $e_1,e_2,e_3$}},\nonumber
\end{eqnarray}
where we have used (\ref{derR}). In the evaluation of this quantity,
the result depends on the order in which the limits are taken. If
one takes the limit $\epsilon\rightarrow 0$ first, one finds that the
diffeomorphism annihilates $R_c$. This means that
$\lim_{\epsilon\rightarrow 0}C_{\epsilon'}(\vec{N})\,M_{\epsilon}(v)=0$.
On the other hand, if we take the limit $\epsilon'\rightarrow 0$
first one gets the result,
\begin{equation}
\lim_{\epsilon'\rightarrow 0}C_{\epsilon'}(\vec{N})\,
R_{c}(M,\epsilon,e)=\left[ (\vec{N}\cdot \nabla M)(v) +
M(v)\lim_{\epsilon'\rightarrow 0}C_{\epsilon'}(\vec{N})\right]
R_{c}(1,\epsilon,e) + O(\epsilon),
\end{equation}
and,
\begin{equation}
\lim_{\epsilon\rightarrow 0}\lim_{\epsilon'\rightarrow 0}C_{\epsilon'}
(\vec{N})\,M_{\epsilon}(v)= (\vec{N}\cdot \nabla M)(v)
\end{equation}
With this choice we see that the action of the diffeomorphism is
what one would have expected geometrically: it Lie-drags the
dependence of $M$ on the position of the vertex. This action
immediately ensures that the action of the diffeomorphism
constraint closes the appropriate commutator algebra on the space
of functions we are considering, provided $\psi(s)$ is either
invariant (like in the case of the Vassiliev invariants) or is such
that the diffeomorphism constraint has the natural geometric
action. In the general case,
\begin{equation}
C(\vec{N}) O(M,\Omega)\psi(s)
= O(N^a \partial_a M,\Omega)
\psi(s) + O(M,\Omega) C(\vec{N}) \psi(s).
\label{diffeoono}
\end{equation}

Based on these calculations for the diffeomorphism, let us now explore
the action of the Hamiltonian constraint on the habitat we have been
discussing. From (\ref{hgen}) and (\ref{derR}) we get
($\epsilon'$ is now the scale parameter of the triangulation
adapted to the spin network),
\begin{eqnarray}
&&H_{\epsilon'}(N)\,R_{c}(M,\epsilon,e) =
{8 \over 3G}
{e}^a_{2}{e}^b_{3} \rho_1(\vec{J}_v)\epsilon_{cdf}
\frac{\epsilon^2}{\epsilon'^2} \int d^3 z
\int_{\bar{e_3}\circ e_2} dy^f  f_{\epsilon'}(z;v)\times\\
&&\left\{ \delta^d_{[b}\partial_{a]z} M(z) \Theta_{\epsilon}(z-y)+
(M(z)+M(y)) \delta(\epsilon-|z-y|)
\frac{\delta^d_{[b}(y-z)_{a]}}{|z-y|}\right\}
+ \mbox{\footnotesize{cyclic permut. in $e_1,e_2,e_3$}},\label{hr}
\nonumber
\end{eqnarray}
where we have written $f_{\epsilon'}(y;v)\equiv
f_{\Box_v}(y)$ in order to make explicit the dependence of the
triangulation with $\epsilon'$. Notice that,

\begin{equation}
\lim_{\{\epsilon,\epsilon'\}\rightarrow 0}
\frac{\epsilon^2}{\epsilon'^2}
\int d^3 z \int_{e} \Theta_{\epsilon}(z-y) f_{\epsilon'}(z;v)=0,
\end{equation}
and,
\begin{equation}
\lim_{\{\epsilon,\epsilon'\}\rightarrow 0}
\frac{\epsilon^2}{\epsilon'^2}
\int d^3 z \int_{e} dy^f \delta(\epsilon-|z-y|)
\frac{(y-z)_{a}}{|z-y|} f_{\epsilon'}(z;v)=0,
\end{equation}
irrespective of the order in which the limits are taken. This
means that,
\begin{equation}
\lim_{\{\epsilon,\epsilon'\}\rightarrow 0}
H_{\epsilon'}(N)\,M_{\epsilon}(v)=0.
\end{equation}
and taking into account (\ref{psiM}),
\begin{equation}\label{[ho]}
H(N)O(M,\Omega) \psi (s) =
O(M,\Omega) H(N) \psi(s).
\end{equation}
This is an important result. It will imply that the Hamiltonian
constraints will commute with each other. We will discuss this in
detail in the next section.

\subsection{The right hand side of the commutator of two Hamiltonians}
In the classical theory, the Poisson bracket of two smeared
Hamiltonian constraints is given by,
\begin{equation}
\left\{H(N),H(M)\right\} = \int d^3x \omega_a(x) q^{ab}(x)
\tilde{C}_b(x)
\end{equation}
with the one-form $\omega_a = (N \partial_a M - M \partial_a N)$
and $\tilde{C}_b(x) =\tilde{E}^b_i(x) F_{ab}^i(x)$ is the
(unsmeared) diffeomorphism constraint. That is, the right hand
side of this expression is a diffeomorphism generated by the vector
obtained by contracting the one-form $\omega_a$ with a
double-contravariant metric. If one wishes this expression to have
a quantum counterpart, it implies realizing the product of the
doubly contravariant metric times a diffeomorphism as a quantum
operator. In order to do this, we need to re-express it in terms of
variables that are suitable for the Ashtekar formulation. This is
done via the following classical identities, as discussed by
Thiemann \cite{QSD3},
\begin{equation}
q^{ab}(x)={1 \over 4} 
\epsilon^{acd} \epsilon_{ijk} \epsilon^{bef} \epsilon_{ilm}
{e^j_c e^k_d \over \sqrt{det(q)}} {e^l_e e^m_f \over \sqrt{det(q)}} 
\end{equation}
and we recall the relationship of the covariant (undensitized) triads with
the more usual contravariant densitized triads,
$\tilde{E}^a_i =\epsilon_{ijk} \epsilon^{abc} e_b^j
e_c^k$. We can now use the key identity, due to Thiemann,
\begin{equation}
2 G  e_a^i(x) =\{A_a^i(x), V\}
\end{equation}
to obtain,
\begin{equation}
q^{ab}(x)={1 \over 16G^2} 
\epsilon^{acd} \epsilon_{ijk} \epsilon^{bef} \epsilon_{ilm}
{\{A_c^j(x),V\} \{A_d^k(x),V\} \over \sqrt{det(q)(x)}} {\{A_e^l(x),V\}
\{A_f^m(x),V\}\over \sqrt{det(q)(x)}}.\label{metc}
\end{equation}

This expression cannot be easily promoted in a direct way 
to an operator due to the denominators involving the determinant of
the metric.  We can achieve this noting that the Poisson bracket of
the connection and the volume depends only on local information of the
volume operator in the surrounding of the point $x$. We can therefore
replace in the above expression the volume of the whole manifold $V$, 
by a ``localized'' volume
(as discussed by Thiemann \cite{QSD3}) $V({\cal{R}}_x)$, given by,
\begin{equation}
V({\cal{R}}_x) = \int d^3 y \,\Theta_{{\cal{R}}_x}(y)
\sqrt{{\rm det}q(y)},
\end{equation}
where ${\cal{R}}_x$ is an infinitesimal closed region around the
point $x$ and $\Theta_{{\cal{R}}_x}(y)$ is the
characteristic function associated with this region. The following
identity holds,
\begin{equation}
\{A_c^i(x),V\} = \{A_c^i(x),V({\cal{R}}_x)\}\,
\label{absorb}
\end{equation}
and in addition, 
\begin{equation}
\lim_{ {\cal{R}} \rightarrow 0} { V({\cal{R}}_x)\over
{\cal V}_{{\cal R}_x} }=
\sqrt{{\rm det}(q)(x)},
\end{equation}
where ${\cal V}_{{\cal{R}}_x}$ is the Euclidean volume of the
infinitesimal region ${\cal R}_x$ given by a fiducial flat metric (we will
see that the final result does not depend on the fiducial metric). 
The above expressions allow us
to absorb the denominators involving the determinant of the metric
into Poisson brackets. Using this results we write (\ref{metc}) in
the form,
\begin{eqnarray}
q^{ab}(x)&=&
{1 \over G^2}
\lim_{{\cal R} \rightarrow 0} \int d^3 z \delta(z-x) {\cal
V}_{{\cal R}_z}\times \\
&&\epsilon^{acd} \epsilon_{ijk} \epsilon^{bef} \epsilon_{ilm}
\{A_c^j(z),\sqrt{V({\cal{R}}_z)}\} \{A_d^k(z),\sqrt{V({\cal{R}}_z)}\}
\{A_e^l(z),\sqrt{V({\cal{R}}_z)}\}\{A_f^m(z),\sqrt{V({\cal{R}}_z)}\}
\nonumber
\end{eqnarray}
We now discretize the integral introducing the triangulation
$\{\Diamond\}=\{\Box_v ,\Delta'\}$ of space defined in section II,
\begin{eqnarray}
q^{ab}(x)&=&{1 \over G^2}
\lim_{{\cal{R}} \rightarrow 0} \lim_{\Diamond\rightarrow 0}
\sum_{\Diamond} {\cal V}_{{\cal R}_v} \Theta_{\Diamond}(x)
\times\\
&& \epsilon^{acd} \epsilon_{ijk} \epsilon^{bef} \epsilon_{ilm}
\{A_c^j(v),\sqrt{V({\cal{R}}_v)}\} \{A_d^k(v),\sqrt{V({\cal{R}}_v)}\}
\{A_e^l(v),\sqrt{V({\cal{R}}_v)}\}\{A_f^m(v),\sqrt{V({\cal{R}}_v)}\}
\nonumber
\end{eqnarray}
where we have identified $v\equiv v_{\Diamond}$ to simplify the
notation. Using (\ref{vol}) we replace the $\epsilon^{acd}$ and
$\epsilon^{def}$ in terms of the volume of the elementary regions
and the edges $u$ of the tetrahedra, and we join four of the $u$'s
with the $A$'s to construct holonomies along the edges of the
triangulation,
\begin{equation}
q^{ab}(x)={1 \over 36G^2}
\lim_{{\cal{R}} \rightarrow 0} \lim_{\Diamond\rightarrow 0}
\sum_{\Diamond} \alpha_{\Diamond}^2\frac{{\cal V}^2_{{\cal{R}}_v}}
{{\cal V}^2_{\Diamond}}\Theta_{\Diamond}(x)
 u^a_p u^b_s\,Q^{ps}_{\Diamond}({\cal{R}}_v)
\label{metc2}
\end{equation}
where we have defined,
\begin{eqnarray}
Q^{ps}_{\Diamond}({\cal{R}}_v)&:=&\epsilon^{pqr} \epsilon^{stn}
\epsilon_{ijk} \epsilon_{ilm}
{\rm Tr}[\tau^j h(u_q)\{h^{-1}(u_q),\sqrt{V({\cal{R}}_v)}\}]
{\rm Tr}[\tau^k h(u_r)\{h^{-1}(u_r),\sqrt{V({\cal{R}}_v)}\}]
\times \nonumber\\
&&{\rm Tr}[\tau^l h(u_t)\{h^{-1}(u_t),\sqrt{V({\cal{R}}_v)}\}]
{\rm Tr}[\tau^m h(u_n)\{h^{-1}(u_n),\sqrt{V({\cal{R}}_v)}\}]\,.
\label{Q}
\end{eqnarray}
We are now ready to promote the above expression as an operator
acting over the spin network wavefunctions, which we assume to be
the loop transform of a state in the connection representation. As in
the case of the Hamiltonian constraint, we adapt the triangulation
to the graph of the spin network choosing the points $\{v\}$ of the
boxes $\{\Box_v\}$ coincident with the vertices of the spin
network, and we choose the regions ${\cal R}_v = \Box'_v$ equal to
that defined by the triangulation around $v$ but with a length
scale $\epsilon'$. With this prescriptions we get from
(\ref{metc2}),
\begin{equation}
{q}^{ab}(x)\,\psi(s)={16 \over 9G^2}
\lim_{\{\epsilon,\epsilon'\} \rightarrow 0}
\sum_{v\in s} \frac{\epsilon'^6}{\epsilon^6}
\Theta_{\epsilon}(x;v)
\int_{e_p} dw^a \Theta_{\epsilon}(y;w)
\int_{e_s} dt^b\,\Theta_{\epsilon}(t;v)
{Q}^{ps}(v)\,\psi(s)\,,
\label{metq}
\end{equation}
where we have used (\ref{tangent}) and $\Theta_{\epsilon}(x;v)
\equiv \Theta_{\Box_v}(x)$ with $\epsilon$ the length scale of the
regions $\Box_v$. The action of the operator ${Q}^{ps}(v)$ on
the spin network wavefunctions is calculated as usual through its
action on the Wilson net appearing in the loop transform. It
is important to notice that this operator looses all dependence
with the scale parameters involved in the triangulation and the
localized volume. This is due to the fact that, acting on the
Wilson net, the holonomies $h(u)$ that appear in the quantum
version of (\ref{Q}) generate (in the limit $\Diamond\rightarrow
0$) finite recoupling coefficients and that the action of the
volume operator is local (it depends only on the vertex included in
the region ${\cal{R}}_v)$.

With this result at hand it is straightforward to
quantize the right hand side of the Poisson bracket of two smeared
Hamiltonian constraints. From the classical expression we
immediately write,
\begin{eqnarray}
\widehat{RHS} \psi(s) &=& \int d^3x \omega_a(x) {q}^{ab}(x)
{C}_b(x)\\
&=& \lim_{\delta\rightarrow 0}
\sum_{e\in s} \int d^3x \int_{e} dy^c
[\omega_a(x) {q}^{ab}(x) +\omega_a(y) {q}^{ab}(y) ]
f_{\delta}(x,y) \Delta_{cb}^{(e)}(\pi_y^x) \psi(s)\,, \nonumber
\end{eqnarray}
where we have used the regularized expression of the unsmeared
diffeomorphism operator (see the companion paper, section II.B).
Introducing now the regularized metric operator we get,
\begin{eqnarray}
\widehat{RHS} \psi(s) &=&
{16 \over 9G^2} \lim_{\{\epsilon,\epsilon',\delta\}\rightarrow 0}
\sum_{e\in s} \sum_{v\in s}\frac{\epsilon'^6}{\epsilon^6}
\int d^3x \int_{e} dy^c
[\omega_a(x) \Theta_{\epsilon}(x;v) +\omega_a(y)
\Theta_{\epsilon}(y;v)] \,f_{\delta}(x,y)\times\nonumber\\
&&\int_{e_p} dw^a \Theta_{\epsilon}(w;v)
\int_{e_s} dt^b\,\Theta_{\epsilon}(t;v)
{Q}^{ps}(v)\Delta_{cb}^{(e)}(\pi_y^x)\, \psi(s)\,.
\end{eqnarray}
At this point it is worthwhile mentioning that this expression has
a regularization ambiguity, given by the two regularization
parameters $\epsilon$ and $\epsilon'$. The latter was introduced, as
we discussed above, in the definition of the localized volume
operator. This ambiguity, associated with the fact that the
localized volume and the ordinary one are the same on these spaces
of functions was first noticed by Lewandowski and discussed in
\cite{GaLeMaPu}. If one chooses $\epsilon=\epsilon'$ one notices
quickly that the expression for the right-hand-side vanishes. The
powers $\epsilon'^6\over \epsilon^6$ cancel and one is left with two
integrals along $w$, $t$ which are of order $\epsilon$ each. The
rest of the expression is simply the action of a diffeomorphism.
Assuming that the latter is finite, the expression therefore
vanishes.

If we trace back the origin of this cancellation, one notices that
the expression (\ref{metq}) for the contravariant metric operator
vanishes identically over spin network states of any kind, in
particular, the Vassiliev invariants we are considering in this
paper. This property is a consequence of the following two
peculiarities of the quantization of $q^{ab}$ in terms of spin
network wavefunctions: the sum over the triangulation reduces to a
sum over the vertices of the spin net (which includes a finite number of
terms), and in each term the result of ${Q}^{ps}(v)\psi(s)$ is
finite. Notice that in the classical result (\ref{metc}) one has an
infinite sum of $Q^{ps}_{\Diamond}({\cal{R}}_v)$, each of which
tends to zero in the limit $\{\Diamond, {\cal R}\}\rightarrow 0$.
This limit gives in general a nonzero result. But in (\ref{metq})
we have instead a finite sum of terms ${Q}^{ps}(v)\psi(s)$,
which are independent of $\epsilon$ and $\epsilon'$. This fact
alters the power counting of the factors $\epsilon$ and $\epsilon'$
in such a way that all the terms tends naturally (i.e. choosing
$\epsilon=\epsilon'$) to zero as the triangulation shrinks to a
point.

That the right-hand-side may vanish was already observed in
\cite{GaLeMaPu}. Could one ``tune'' the limits in $\epsilon$ and
$\epsilon'$ so this quantity is nonvanishing? One could, but the
result would be dependent on the background structures introduced
to regularize. If one made the expression non-vanishing it would be
proportional to the normalized tangent vectors at the intersection,
which are background dependent. This is not surprising: there is no
naturally defined second order symmetric contravariant tensor
defined in a manifold without metric. This result is quite strong.
It implies that the commutator of two Hamiltonians will have to
vanish if one wishes to have consistency. We anticipated that this
would happen in the previous subsection and we will now see in
detail how it happens.

As a final remark, we notice that if one wished to define the doubly
covariant metric, it is straightforward to compute it using the
identities introduced by Thiemann and one finds that regularized with
the same procedures we followed up to now, it diverges. Should one
worry about a theory of quantum gravity where the doubly contravariant
metric vanishes and the doubly covariant metric diverges? We will
return to this in the discussion section.

\section{Consistency of the constraint algebra: habitats}

\subsection{Habitat I: holonomies of smooth connections}

The loop derivative was originally introduced in the context of
Yang--Mills theories, where the natural functions to act upon were
holonomies of smooth connections. These functions are not the most
natural ones to consider in the context of diffeomorphism invariant
theories like general relativity, but our constraints are well defined
on them, so we can consider them to be a ``habitat'' where to test the
constraint algebra, at least as a mathematical exercise. On these
kinds of functions the diffeomorphism constraints that we have
introduced here are known to close the appropriate algebra in the
limit in which regulators are removed (see \cite{GaGaPu}, although the
regularization is slightly different than the one we use in this paper,
it is immediate to check that the same calculations go through). We
therefore will not repeat the calculation here. The Hamiltonian
constraint vanishes identically on this habitat. Starting from
equation (\ref{hgen}), if the loop derivative is a smooth function,
one is left with two one-dimensional integrals along the edges of a
cell of a finite function. In the limit in which the triangulation is
shrank, the result vanishes. The commutator of a diffeomorphism with
a Hamiltonian therefore immediately reproduces the classical result.

The right-hand side of the commutator of two Hamiltonians that we
introduced in the previous subsection also vanishes, here one has
three one dimensional integrals along the edges of a triangulation of
a quantity that goes as ${\epsilon'}^6/\epsilon^6$. The final result
therefore goes as ${\epsilon'}^6/\epsilon^3$. Therefore, as long as one
chooses $\epsilon'$ shrinking to zero faster than $\sqrt{\epsilon}$, the
right hand side vanishes.

On this space therefore, we reproduce the classical Poisson algebra at
the level of the commutators. The diffeomorphisms close among
themselves and the Hamiltonian vanishes identically. Could one simply
claim that this is the ``right'' habitat to do quantum gravity?  The
answer is that this is unlikely. Although all states are solutions of
the Hamiltonian constraint, this space does not contain any solution
to the diffeomorphism constraint. Solutions to the diffeomorphism
constraint in terms of holonomies can only be constructed as infinite
superpositions, functional integrals or ``group averaging'', and in
these cases one includes connections that are not smooth. 

An interesting point is that in this habitat, since one does not have
solutions of the diffeomorphism constraint but has solutions to the
Hamiltonian constraint, the only way that one could achieve
consistency in the algebra is if the right hand side of the commutator
of two Hamiltonians vanishes, which is the case.

Another point to consider is that if one examines the expression of
the commutator of two Hamiltonians, although both members vanish in
the limit in which one shrinks the triangulation ---given the
smoothness of the loop derivative--- away from the limit the calculation
is problematic. For instance, it is not clear what is the action of
the volume operator on a state given by the loop derivative of a
smooth holonomy. One might consider introducing a definition of the
volume on this kind of space, but this has yet to be done in detail.

\subsection{Habitat II: The framing-independent Vassiliev invariants}

Remarkably, this space of invariants also leads to a reproduction of
the classical Poisson algebra at a trivial level only (as usual, we
limit our discussion to trivalent intersections). This is based
on the fact proved in the appendix of the previous paper, that these
invariants are annihilated by the loop derivative that appears in both
the diffeomorphism and the Hamiltonian constraints and of the
expression of the RHS. Therefore the ambient isotopic
(framing-independent) Vassiliev invariants for spin networks with
trivalent intersections are annihilated by all the constraints of
quantum gravity and consistently, by the right hand side of the
commutator of two Hamiltonian constraints. It should be pointed out
that the space is in no way trivial: as we discussed in the appendix
of the previous paper, the annihilation is a detailed property of the
space of Vassiliev invariants, related to the decomposition of the
invariants in framing independent and framing dependent components and
the detailed structure  
of chord diagrams appearing in the coefficients. Hints that
the Vassiliev invariants for trivalent intersections were annihilated
by the Hamiltonian constraint were found for the lower invariants in
terms of loops \cite{BrGaPuprl},\cite{Gr}, in the lattice
\cite{GaPurigor}. 

One could mention as a more trivial example of a space with similar
property the states based on diffeomorphism invariants of 
spin networks with no vertices, which are also trivially
annihilated by all the constraints we consider. This is an extension
of results also first suggested in the context of loops \cite{RoSm}.

It should be remarked that this property of the loop derivative is not
true for higher valence intersections. Future extensions of the
Hamiltonian constraint to higher valence intersections could be tested
in this habitat for non-trivial consistency.

\subsection{Habitat III: Vassiliev invariants}

By this space we mean the invariants that appear in the power series
expansion of the expectation value of the Wilson net in a
Chern--Simons theory in terms of the inverse coupling constant
$\kappa$. In particular, the whole series is an example of such an
invariant. Each coefficient of the series and certain portions of them
are also examples. In general, these invariants are sums and products of the
independent invariants that appear at each order, both
framing-independent and framing-dependent (see previous paper). These
states have the important property that the loop derivative that
appears in the expressions of the constraints (which is evaluated on a
path $\pi$ of infinitesimal length) can be rewritten simply as,
\begin{equation}
\Delta_{ab}(\pi_x^y) V(s) =\sum_{e_x \in s} \int_{e_x} dz^c \epsilon_{abc}
\delta^3(z-y) V'(s)
\end{equation}
where $e_x$ is the edge of the spin net containing the point $x$ and
where the invariant $V'(s)$ is another Vassiliev invariant of one
order less than $V(s)$. What happens is that because of the
infinitesimal length of $\pi$ one can rearrange the action of the loop
derivative in terms of the original spin network $s$ using recoupling
identities, at the expense of some additional group factors, which we
reabsorb notationally in $V'(s)$. We will see that we do not need the
details of the relation of $V(s)$ to $V'(s)$ for proving the
consistency. As usual, all our discussion is limited to trivalent
intersections. The relation above implies that for the Vassiliev
invariants, one can write,
\begin{equation}
H(N) V_n(s)= 
-{1 \over 3G}
\sum_{v\in s} N(v)  \nu_{\vec{J}_v}
V_{n-1}(s).
\end{equation}
This can be seen by considering equation (\ref{nu}) and recalling that
$E(s,\kappa)=\sum_{n=0}^\infty V_n(s) \kappa^n$.  In particular, we
see that the action of the Hamiltonian on these states can be written
(up to a constant factor which we will omit for simplicity) as,
\begin{equation}
H(N) V_n(s) = {O}(N,\nu) V_{n-1}(s). \label{honvas}
\end{equation}

\subsubsection{Commutator of two diffeomorphisms}

On this space, the diffeomorphism constraint vanishes identically, as
we discussed in the previous paper. So the algebra of diffeomorphisms
is trivially satisfied. Nontrivial commutators to be realized will be
those of a diffeomorphism with a Hamiltonian and that of two
Hamiltonians. 

\subsubsection{Commutator of a diffeomorphism with the Hamiltonian}

In the case of a diffeomorphism with a Hamiltonian, since the
wavefunctions are diffeomorphism invariant, of the two terms of the
commutator one is left with the one in which the diffeomorphism acts
at the left. Since the action of the Hamiltonian on $V(s)$ is not
diffeomorphism invariant, one should recover the proper action of the
diffeomorphism through such term. Here the derivations of section
\ref{marked} become useful. We start from the expression
(\ref{honvas}), and then use (\ref{diffeoono}) to get,
\begin{equation}
C(\vec{N}) H(M)V_n(s) 
= {O}(N^a \partial_a M,\nu)
V_{n-1}(s) + O(M,\nu) C(\vec{N}) V_{n-1}(s)
= H(N^a \partial_a M)
V_{n}(s). 
\end{equation}
and from here the correct commutator follows immediately. This implies
that the quantum Hamiltonian transforms covariantly in a correct way. This
calculation is one of the main differences of our construction with
respect to the one of Thiemann \cite{QSD1}, since in that context
one does not consider an infinitesimal generator of diffeomorphisms.

\subsubsection{Commutator of two Hamiltonians}

As we discussed in section \ref{marked}, the action of a Hamiltonian
on a function with marked points as one gets after acting with a
Hamiltonian on a Vassiliev invariant was given by expression
(\ref{[ho]}), which we can combine with 
the action of the Hamiltonian (\ref{honvas}) to get,
\begin{equation}
H(N) H(M) V_n(s)=H(N) {O}(M,\nu) V_{n-1}(s)=
{O}(M,\nu) H(N) V_{n-1}(s)={O}(M,\nu) {O}(N,\nu)V_{n-2}(s).
\end{equation}
And if one performs the calculation in the reverse order, one obtains
the ${O}$ operators in the reverse order. However, these operators
are multiplicative, so they commute. Therefore one has the result,
\begin{equation}
[H(N),H(M)] V_n(s)=0,
\end{equation}
which is the expected commutation relation on states that are invariant
under diffeomorphisms.

\subsection{Habitat IV: diffeomorphism dependent functions}

The states we have already considered, where one has a Vassiliev
invariant times a scalar function that depends on the position of the
vertices of the spin network, are an attractive habitat where
the diffeomorphism constraint does not vanish.

\subsubsection{Commutator of diffeomorphisms}

As discussed in section (\ref{marked}) the diffeomorphisms on these
states reduce to Lie dragging of the scalar functions of the vertices
(see formula (\ref{diffeoono})). Since they correspond to a natural
geometric action, the consistency of the algebra of diffeomorphisms is
immediate.

\subsubsection{Commutator of diffeomorphism with a Hamiltonian}

We start from the action of a Hamiltonian on these kinds of states, 
\begin{equation}
H(N) O(M,\Omega) V_n(s) = O(M,\Omega) O(N,\nu) V_{n-1}(s),
\end{equation}
where we have used the fact that $O$ and $H$ commute and that the action
of the Hamiltonian on these states produces a Vassiliev invariant of order
lowered by one unit times a vertex function. We now act with a diffeomorphism,
\begin{equation}
C(\vec{L})H(N) O(M,\Omega) V_n = C(\vec{L})O(M,\Omega) O(N,\nu) V_{n-1}(s),
\end{equation}
and now use the fact that Leibniz' rule applies to the action of the
diffeomorphism constraint (stemming from the fact that it also applies to the action of the loop derivative),
\begin{equation}
C(\vec{L})O(M,\Omega) O(N,\nu) V_{n-1}(s) = 
O(L^a \partial_a M,\Omega) O(N,\nu) V_{n-1}+
O(M,\Omega) O(L^a \partial_a N,\nu) V_{n-1},
\end{equation}
and we now reconstruct the Hamiltonian in the first and second terms,
\begin{eqnarray}
O(L^a \partial_a M,\Omega) O(N,\nu) V_{n-1}+
O(M,\Omega) O(L^a \partial_a N,\nu) V_{n-1}&=&
O(L^a \partial_a M,\Omega) H(N) V_{n}\\&&+
O(M,\Omega) H(L^a \partial_a N) V_{n},\nonumber
\end{eqnarray}
and using the fact that $H$ and $O$ commute, we get,
\begin{equation}
C(\vec{L})H(N) O(M,\Omega) V_n  = H(N) O(L^a \partial_a M,\Omega)  V_n
+H(L^a \partial_a N) O(M,\Omega) V_{n},
\end{equation}
which we can rewrite as,
\begin{equation}
C(\vec{L})H(N) O(M,\Omega) V_n  = H(N) C(\vec{L})O(M,\Omega)  V_n
+H(L^a \partial_a N) O(M,\Omega) V_{n},
\end{equation}
from where we get the correct commutation relation,
\begin{equation}
[C(\vec{L}),H(N)] O(M,\Omega) V_n  = 
H(L^a \partial_a N) O(M,\Omega) V_{n}.
\end{equation}
\subsubsection{Commutator of two Hamiltonians}
This calculation proceeds along the same lines as the one in the previous
subsection, we essentially rewrite the action of the Hamiltonian in terms
of the $O$ operator, and note that the $O$ operators commute, 
\begin{equation}
H(L)H(N) O(M,\Omega) V_n(s) = O(L,\nu) O(N,\nu) O(M,\Omega) V_{n-2}(s) =
H(N) H(L) O(M,\Omega) V_n(s),
\end{equation}
and we therefore get that $[H(N),H(M)]=0$, which as we discussed before, is
consistent with the representation in these spaces of functions of the 
right hand side of the quantum commutator.

\section{Discussion}
\subsection{Summary}
We have presented a canonical quantization of the constraints of
canonical general relativity. We represented the diffeomorphism and
Hamiltonian constraints using two novel ingredients: the loop
derivative to represent the field tensor $F_{ab}$ and the use of
spaces related to the generalization of the Vassiliev invariants to
spin networks (restricted to trivalent intersections) as
wavefunctions. In terms of the latter we constructed several
``habitats'', including spaces of functions that are not invariant
under diffeomorphisms and we checked that one obtained a consistent
algebra of quantum commutators of the constraints. Consistency in this
sense implies that the quantization of the canonical Poisson
identities between the classical constraints is implemented correctly
in the quantization. We have observed that this consistency is
achieved at the price of having a vanishing right hand-side for the
commutator of two Hamiltonians and that we can trace back this fact to
the vanishing nature of the doubly covariant metric operator in these
spaces of functions.

\subsection{Comparison with Thiemann's results}

We should point out common elements and differences with the
quantization presented by Thiemann \cite{QSD1}. In Thiemann's case the
Hamiltonian was implemented on the space of diffeomorphism invariant
cylindrical functions of spin networks. On these spaces one does not
have a well defined notion of the field tensor $F_{ab}$ and the
functions are not loop-differentiable. In our case, the availability
of the loop derivatives allows to have at hand a ``differential
calculus'' that allows for several novel constructions. Examples of
them are the calculations performed in computing the constraint
algebra, and the possibility of finding novel solutions to the
Hamiltonian constraint, based on the behavior of the Vassiliev
invariants under loop differentiation.

Another difference involves the implementation of an infinitesimal
generator of diffeomorphisms. In Thiemann's original construction one
worked directly in terms of diffeomorphism invariant states and
therefore one did not have an infinitesimal generator.  The
construction was extended by Lewandowski and Marolf \cite{LeMa} to
``habitats'' that are dependent on diffeomorphisms. With that
extension, Thiemann's work achieves the same level of consistency that
we have in this paper, in the sense that one can check non-trivial
commutators of diffeomorphisms and diffeomorphisms with
Hamiltonians. One still has the feature of a vanishing right hand side
of the commutator of two Hamiltonians \cite{GaLeMaPu}. As we discussed
in this paper, this feature appears as inescapable in the context of
wavefunctions only dependent on spin networks as the ones we
considered here, since as we mentioned one does not have enough
structures to construct a naturally defined symmetric metric tensor.

Other differences arise insofar as the space of wavefunctions
considered. In the case of Thiemann, one had spaces of functions with
well defined inner products, which allow to discuss normalizability
and study the spaces of solutions with a level of rigor that is not
available currently in our approach, since we do not have an inner
product on the habitats we are considering. It might not be impossible
to find a suitable inner product with the same techniques that led to
the construction of the measures on the spaces of cylindrical
functions, but it has not been achieved yet. Concerning solutions of
the constraints, one has some available both in our approach and in
Thiemann's, that appear as quite distinct in their features. In
Thiemann's approach, the solutions are obtained using group averaging
techniques. This leads to structures like ``tassels'' and others
\cite{tassles} in which accumulation of lines and vertices take
place. In our approach, one has a number of solutions of the
Hamiltonian constraint (the framing independent Vassiliev invariants
for trivalent intersections) that do not depend on such structures. On
the other hand, they appear as ``trivial'' solutions in the sense that
they may not  exist if one considers intersections of higher
valences. In our approach one may also construct solutions to the
Hamiltonian constraint with a cosmological constant, following similar
ideas that led to the construction of states in terms of loops
\cite{BrGaPunpb}. Although we have not pursued this in detail yet, it
appears quite plausible that these types of solutions exist, given the
structure of the extra term due to the cosmological constant in the
Hamiltonian constraint in terms of the spin network  approach
\cite{GaGrPu98}. 

Thiemann's approach has also been studied in $2+1$ dimensions
\cite{QSD4}, and appears to lead to a satisfactory quantization,
provided one chooses in an ad-hoc way an inner product that rules out
certain infinite dimensional set of solutions. In a forthcoming paper
we will discuss the quantization of $2+1$ dimensional gravity using an
approach that has elements in common with the one we pursue here, in
particular the requirement of loop differentiability of the states.
We will see that this requirement limits us (at least for
low valence intersections) to the correct solution space in a natural
way.

\subsection{Is the theory satisfactory?}

The reader might find unsatisfactory that the right hand side of the
commutator of two Hamiltonians vanishes. Even more unsatisfactory may
appear the fact that this is due to the vanishing of the doubly
covariant metric in this approach to quantum gravity. In this
subsection we will address this and other issues.

The first question that one may raise is if this is not just a
pathology that is introduced by our limitation to trivalent
intersections for reasons of calculational convenience. After all, one
knows that this subspace in in some sense degenerate since the volume
operator vanishes identically. Unfortunately, it appears unlikely that
extending our results to higher valent intersections (to which we do
not see any technical obstruction, apart from greater calculational
complexity) will change things insofar as the commutator of two
Hamiltonians. The action of the Hamiltonian constraint on higher
valent intersections is more complicated largely because the volume
operator and the action of the loop derivative cannot be reduced to a
simple group-dependent prefactor times the original state, but will in
general involve a linear combination of states, with a non-trivial
group-dependent matrix of coefficients. However, the dependence of the
Hamiltonian on the smearing function will remain as a multiplicative
one. Therefore it appears that the commutator of two Hamiltonians will
again vanish. Moreover, the reasons we gave for the vanishing of the
right hand side of the commutator are independent of the valences of
the intersections of the spin networks. The expression for the
doubly-covariant metric will be more complicated, but will still
include a prefactor with the same dependence on the regulators as in
the trivalent case and it will vanish. One therefore expects
consistency at the same level we achieved for the trivalent case.

Can one consider satisfactory a theory with a vanishing doubly
contravariant metric (and a divergent doubly covariant metric)?  The
answer to this will only appear in a definitive way when one
constructs physical predictions from the theory, which at the moment
are lacking. One can get partial indications that the pathology might
not be as severe as it appears at first sight from the fact that one
can define reasonable quantities like areas, volumes and lengths in
this framework in spite of having the ill defined metric operators. An
attractive feature of the Vassiliev states is that other
diffeomorphism invariant operators can be defined as well. For
instance, if one considers the integral $\int \sqrt{g} {\rm Tr}(F_{ab}
F^{ab})$, given that the Chern--Simons states are such that the
densitized triads are proportional to the magnetic vector constructed
from the curvature, one finds that quantum representation of this
integral is identical to the volume operator.

An important observation concerning how satisfatory is a theory with a
vanishing metric operator like the one we propose is that in
$2+1$-dimensional gravity similar pathologies appear, yet the correct
physical theory is recovered. If one pursues a quantization similar to
Thiemann's \footnote{In a forthcoming paper we will present a
quantization of $2+1$ dimensional gravity along the same lines as the
one proposed in this paper, and the observations we make in this
subsection apply to it as well. We will also expand the discussion on
the physical observables and their connection with the kinematical
metric we present in this subsection.} in $2+1$=dimensions (this is
discussed in detail in \cite{QSD4}) one notices that the Hamiltonians
also commute. Moreover, the geometrical arguments leading to the
vanishing of the $q^{ab}$ operator still hold in this context: there
is no quantity one can build out of loop states that will yield a
doubly contravariant symmetric tensor. Yet, it was shown in
\cite{QSD4} that one can recover the correct physical theory.
How can these apparently contradictory elements be reconciled? It has
to do with the nature of the kinematical calculations we are
performing. To check the ``off shell'' contraint algrebra one is
required to operate with a kinematial space of wavefunctions that are
not annihilated by the constraints. When one recovers ``physics'' one
should really do it with states that are annihilated by the
constraint. This can be carried out in detail in the $2+1$ dimensional
case. Suppose one wishes to ask questions about the ``metric'' of the
theory. The first observation is that on physical states one cannot
define a metric tensor since the latter states are diffeomorphism
invariant and the metric tensor is not. One could gauge fix a metric
tensor, say by asking questions about the value of its components in a
fixed coordinate system. These questions can be answered: the
components of the metric in a fixed coordinate system can be related
to the values of the invariant operators $T^0$ and $T^1$ in such a
coordinate system (and therefore in general, since the latter are
coordinate invariant). One therefore has well defined operators
associated with the metric that can be evaluated in the physical space
of states. But such operators have little to do with the $q^{ab}$ we
introduced in the kinematical space. It is not obvious at all that the
two operators will be related in any way since the former are non-vanishing
whereas the latter vanishes identically.

There is a certain disquieting element in the last observation, since
it seems to imply that one should be careful before drawing
conclusions from calculations at a kinematical level. This in
particular, implies all calculations involving the constraint
algebra. When we set out to work on the current paper, our expectation
was that reproducing the contraint algebra  would be a strong test for
our quantization procedure, that perhaps would rule out all but a few
of the possible theories. What we have learned is that this is
probably not the correct view on the issue: one can obtain consistency
in many ways at a kinematical level (in particular with $q^{ab}=0$).
One can construct many consistent theories of quantum gravity. And
still, what the $2+1$ dimensional example shows is that the
``physics'' of all these theories is really deeply buried in the
states that solve the constraints. In a sense this is good, since the
theory we are proposing here differs significantly at this level from
the one proposed by Thiemann and therefore further adds to our options
for trying to match experimental results. 

The observation that kinematical calculations may have quite
nontrivial connections with ``real physics'' poses problems for
computations that try to obtain heuristic physical insights from
looking at properties of the kinematical states (the ``weave''
approach to the semiclassical theory). An example of this are our own
results on gamma-ray-burst light dispersion \cite{GaPu}. From the
$2+1$ example we see that quantum fluctuations in the ``physical
metric'' (the ones one would expect to influence the propagation of
matter fields) may not have a direct connection with fluctuations of
the kinematical metric. Yet it is the latter that are used in the
concrete computations of \cite{GaPu}.

Concerning the latter point, it is worthwhile mentioning explicitly 
that the theory presented in this paper 
can be coupled  to matter and have well defined
expressions for the Hamiltonians of matter fields, which also involve
the metric in a non-trivial way (the coupling of the theory to matter
can, for instance, be achieved using the same construction as Thiemann
\cite{QSD3}; or one could consider alternative settings in which the
loop derivative is also used in the representation of certain matter
fields). It is somewhat unfortunate that the remarkable results on
black hole entropy that have been achieved within this context
\cite{AsBaCoKr} do not involve in a detailed way the action of the
Hamiltonian constraint in the ``bulk'' of the spacetime to be used as
guideline for constructing the constraints, and therefore cannot
distinguish at the moment our proposal from Thiemann's.  

The vanishing of the doubly contravariant metric tensor, apart from
appearing as a ``robust'' feature based only on elementary notions of
covariance of the elements involved in constructing the spin network
states, is the way in which our approach bypasses the ``hermiticity
problem'' of the canonical quantization. As we mentioned, if one were
to demand that the diffeomorphism and Hamiltonian constraints and the
doubly covariant metric be Hermitian operators (since they correspond
to real classical quantities), the commutator of two Hamiltonians
(which is Hermitian) cannot be simply equal to the product of a metric
times a diffeomorphism (which is not Hermitian). One could fix the
Hermiticity of the right hand side by ``symmetrizing'' the operator,
but then one could face an anomaly problem, since the diffeomorphism
constraint would not act to the right of the product. This difficulty
is bypassed (in the context of non-diffeomorphism invariant states) if
the metric vanishes. It should be noticed that it is not obvious that
one should promote the constraints to Hermitian operators, for
instance, see \cite{MaGi} for counterexamples in the context of 
the quantization of non-unimodular gauge groups.

\subsection{Is the theory unique?}

\begin{figure}
\centerline{\psfig{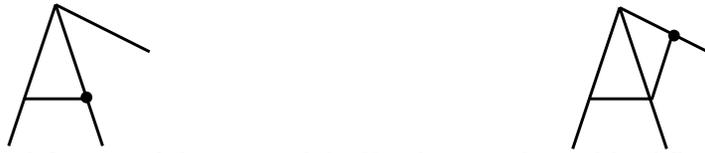}}
\caption{Possible alternative definitions of the action of the
Hamiltonian, obtained by different choices of path in the loop 
derivative.}\label{fifi}
\end{figure}

Related with the issue of the lack of guidelines to construct the
theory is the problem of uniqueness. We already have a manifest
non-uniqueness in that the theory presented here and the theory
introduced by Thiemann appear as both satisfactory yet distinct.  The
ambiguity is worse than this. Even if one stays within one general
approach, for instance considering the space of Vassiliev invariants
as ``arena'' for quantization, there are many ways in which one could
implement the Hamiltonian constraint. We have not analyzed them in any
detail, but some salient features of them are worthwhile
mentioning. The ambiguities arise in the various limits involved in
constructing the Hamiltonian. One of them is associated with the
specific action of the loop derivative. The loop derivative is
dependent on a path. In this paper we have taken such a path to
coincide with one of the lines of the spin network. We then acted with
the loop derivative and collapsed its action using recoupling
identities. The final action of the Hamiltonian was therefore
``ultra-local'' in the sense that it returned a wavefunction with the
same vertex structure times a vertex dependent prefactor (for valences
higher than three one gets a linear combination with different spin
weights for the intertwiners yet the topology of the vertex is
unchanged). This may raise concerns that ``super-selections'' could
appear in the sense that it could be easy to construct operators that
commute with the Hamiltonian. One could define a different Hamiltonian
in a straightforward way, by assuming a different topology for the
path associated with the loop derivative.  For instance, one could
assume that the path starts at the vertex, advances along one of the
edges of the spin network and then crosses towards another edge of the
spin network along one of the edges of the tetrahedra introduced in
the discretization. At the end of the path, the loop derivative
acts. One then is left with an action of the Hamiltonian resembling
the one introduced by Thiemann: the constraint produces a vertex
dependent prefactor, but it also alters the structure of the spin
network by ``adding a line'' at the vertex. A different proposal would
be to have a path that starts at the vertex, advances along one of the
edges of the spin network, crosses towards another edge along the
tetrahedron and then continues to cross towards another edge of the
spin network, as shown in figure \ref{fifi}.

In the first of these proposals (as in Thiemann's), the 
Hamiltonian adds vertices called ``exceptional'' in the sense that
they are planar vertices. The Hamiltonian vanishes identically on such
vertices, because the volume vanishes on planar vertices, even if they
are four-valent. In the second proposal, the intermediate vertex that
is created is non-planar. This might raise hopes that the commutator
of two Hamiltonians may be less trivial, since the second Hamiltonian
acts at the newly created vertex, but more careful analyses seem to
indicate that the Hamiltonians still commute \cite{GaLeMaPu}. If it
they were not to commute one would still be faced with how to
reproduce a similar vertex structure with the right hand side operator
in a natural way. Summarizing this portion of the discussion: it
appears that there is a non-trivial, possibly infinite, 
amount of ambiguity in the definition of the theory, that is not 
significantly constrained by the imposition of the correct quantum
commutator algebra.

Related to the latter point is the fact that due to the vanishing of
the metric, the consistency check provided by the commutator of two
Hamiltonians is less detailed than if one had to prove the equality of
non-vanishing operators. We already see that we have two distinct and
apparently consistent, quantizations of the constraints (with several
possible variants of each). It will require further study to determine
if one quantization is ``better'' in the sense of reproducing expected
results than the other. One possibility would be to consider the
commutation of the Hamiltonian with various operators, and study if
inconsistencies appear. This procedure, however, might be limited by
the fact that most commutators have non-trivial right hand sides which
will inevitably involve quite a bit of ambiguity at the time of their
quantization.

Another point that might be raised is that these approaches appear
confined to four dimensions and to the Einstein-Hilbert action, and
therefore one may have little hopes of making contact with other
approaches, as those based on string theories. These approaches not
only may present avenues to understand quantum gravity but they also
have the attractive feature of unifying gravity with other
interactions, a goal some may consider desirable in itself. It should
be pointed out however, that progress is being made \cite{FrKrPu}
towards describing $N$ dimensional general relativity in terms of
connections. Although constructions like the ones we discussed here
have not been pursued in detail in this context, they appear as
plausible.

\subsection{Final remarks}

To conclude, we have at the moment canonical quantizations of general
relativity (possibly coupled to matter) that appear as mathematically
consistent at the kinematical level at which they have been
studied. This level of consistency had never been achieved before in
other approaches. Further exploration of the consequences of the
quantization will be needed to determine if any of them are physically
satisfactory theories of the quantum gravitational field. In
particular the exploration of the space of states that solve the
constraints and how physical quantities can be evaluated on them
appears as a natural next step in the quantization program.

\acknowledgements We wish to thank Abhay Ashtekar, Laurent Freidel,
John Klauder, Karel Kucha\v{r} and  Thomas Thiemann for comments and
discussions.  This work was supported in part by the National Science
Foundation under grants NSF-PHY-9423950, NSF-INT-9811610,
NSF-PHY-9407194, research funds of the Pennsylvania State University,
the Eberly Family research fund at PSU.  JP acknowledges support of
the Alfred P. Sloan and John Simon Guggenheim foundations. We
acknowledge support of PEDECIBA (Uruguay). RG and JP wish to thank the
 Institute for Theoretical Physics of the University of California at
 Santa Barbara and CDB, RG and JG the Center for Gravitational Physics
 and Geometry at Penn State for hospitality during the completion of
 this work.

\end{document}